# Strain-induced structural transformations in hexagonal monolayer boron nitride. A density functional study

O. Bystrenko[1], B. Ilkiv, S. Petrovska, O. Foia, T. Bystrenko, O. Khyzhun

*Frantsevich Institute for Problems of Materials Science,*
*National Academy of Science of Ukraine, Kyiv, Ukraine*

**Abstract:** In this work we theoretically study the structural transformations occurring in monolayer hexagonal boron nitride (h-BN) subjected to strong deformations. The first-principle computer simulations are based on the density functional theory employing the pseudopotentials with the exchange-correlation energy given in Perdew-Burke-Ernzerhof form. Depending on the direction and the type of the strain, the simulations reveal different types of structural transformations in two-dimensional (2D) h-BN associated with the elastic failure. It is found that, under certain shear strains, there can form an ordered 2D orthorhombic phase with the 2D layer space symmetry *pmmm*. Thermal stability of the latter up to ~900 K is demonstrated by first-principle Born-Oppenheimer molecular dynamics. The examination of electronic structure indicates that, in contrast to the conventional h-BN, which is an insulator with the wide band gap, the predicted orthorhombic phase is expected to be a semiconductor with the narrow band gap of ~ 0.65 eV.

**Keywords:** monolayer boron nitride, phase transformation, strain, DFT

## I Introduction

In recent years, since the discovery of graphene (Novoselov et al. 2004), two-dimensional materials attracted considerable attention of researchers. Low-dimensional crystal systems have unique optical, electronic and mechanical properties that make them very promising for the use in micro- and optoelectronics.

Effects produced by mechanical deformations, like stretching or bending, on the electronic structure and optical characteristics of 2D materials, are of particular interest in the context of their possible applications in tunable flexible devices. Well known examples are the strain-induced changes in electronic properties and structural or semiconductor-metal transitions in transition metal dichalcogenides (Kang and Kwon 2019; Liu 2023)

In this work we examine structural transformations occurring in monolayer hexagonal boron nitride (2D h-BN) subjected to strong tensile and shear deformations. The basic mechanical characteristics of 2D h-BN (like elastic moduli, tensile strength, etc.), as well as the dependence of its electronic properties on elastic strains are currently relatively well understood (see, for instance, (Kürkçü and Yamçıçıer 2019; Satawara et al. 2021; Wardah et al. 2019; Bhattacharya et al. 2012; Topsakal et al. 2009)). At the same time, the strain-induced structural transformations occurring outside the elastic regime remain poorly known. The study is based on first-principle density functional theory (DFT) which has proven to be an accurate and reliable tool for solving such problems.

[1] Corresponding author. E-mail: o.bystrenko@ipms.kyiv.ua

## II Computational methods

In this paper first-principle calculations were carried out by using Quantum Espresso software (Giannozzi et al. 2009). We employed the standard solid state pseudo-potential library (SSSP: precision, version 2.0 (Pradini et al. 2021)) with the exchange-correlation energy given in Perdew–Burke–Ernzerhof (PBE) form (Perdew et al. 1996). The cutoff energy for the basis wave functions and charge were set to 60 and 480 Ry, respectively. The total energy convergence threshold for the electron wave functions was chosen $10^{-7}$ Ry. The number of atoms in supercell was 32 and 16 for calculations of tensile and shear deformations and for first-principle molecular dynamics (MD) simulations, with the number of ***k*-**points in Brillouin zone being 3x3x1 and 6x3x1, respectively.

High precision calculations of phonon dispersion curves for the theoretically predicted orthorhombic phase have been performed for the unit cell containing 4 atoms. The convergence energy threshold and ***k*-**points in Brillouin zone for this case were set as $10^{-14}$ Ry and 16x10x1.

Initial information for h-BN structure was taken from the materials database (Jain et al. 2013). The configuration for two-dimensional h-BN was obtained by removing redundant atomic layers in bulk hexagonal boron nitride. Optimization of the initial structure with respect to ion positions and lattice parameters for free cell geometry in XY dimensions (while keeping Z-dimension c=20 A constant) was performed by means of BFGS method (Billeter et al. 2003).

## III Results of simulations

In this work, we examine the properties of a single 2D h-BN monoatomic layer under the uniaxial tensile and shear strains. The simulations for the case of uniaxial deformations were carried out in the following way. Initially, the carefully relaxed rectangular h-BN structure consisting of 32 atoms was prepared. Then, it was subjected to a series of successively increasing tensile strains in a selected direction. We considered here the zigzag and armchair directions, which we refer to hereafter as X- and Y- strains, respectively.

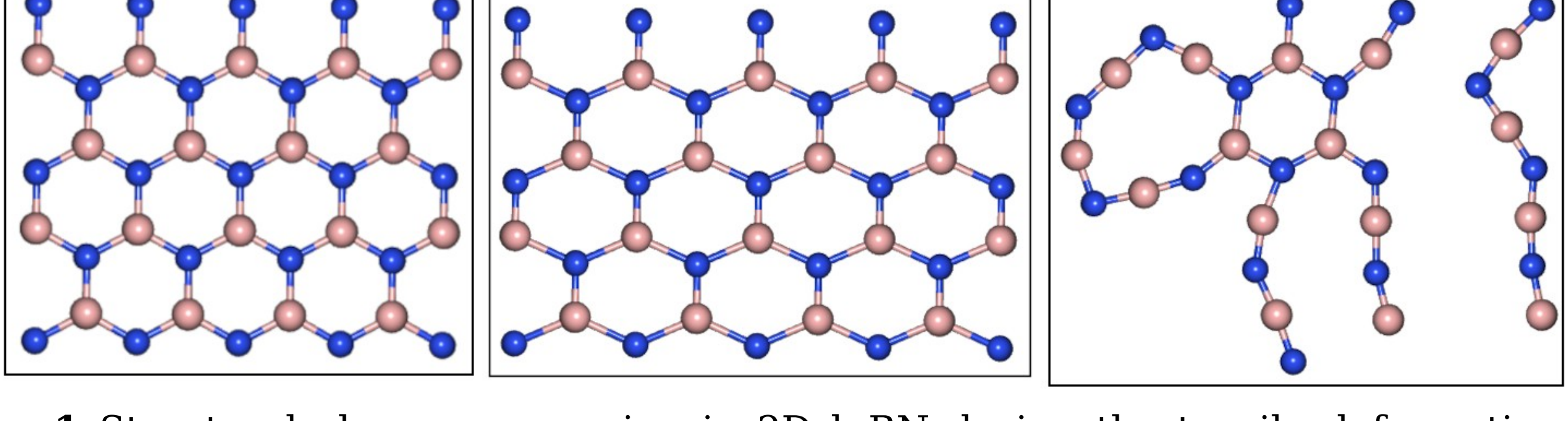

**Fig. 1** Structural changes occurring in 2D h-BN during the tensile deformation in zigzag direction. Left: the initial h-BN lattice; center: relaxed structure at the strain $\Delta x/x = 0.3$; right: relaxed structure at the strain $\Delta x/x = 0.5$.

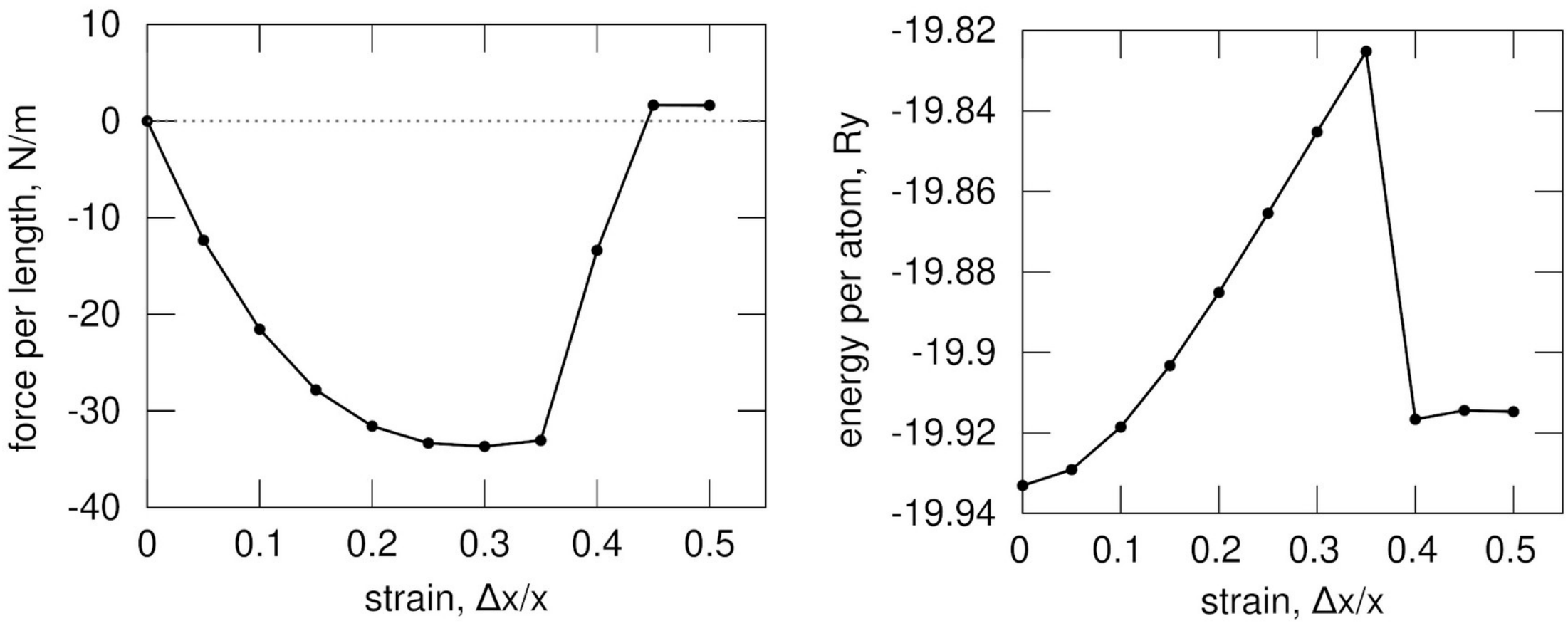

**Fig. 2** Calculated tensile force per length (left) and energy per atom (right) vs strain in X (zigzag) direction for h-BN monolayer.

At each step, the ionic and partial cell relaxation was performed for the fixed deformation in the strain direction, while setting the perpendicular dimension (i.e., Y for X-strains, or X for Y-strains) free. The Z-dimension remained constant in all simulations.

We monitored the behavior of the total energy, tensile force and structural changes occurring during deformations. The results for X-strains are given in Figs. 1 and 2. The elastic regime in this case is observed up to the deformation of $\Delta x/x \sim 0.35$. Larger deformations, ~ 0.35-0.5, result in the lattice destruction, accompanied by the formation of a broken disordered structure. A sharp change in the behavior of force and energy at the deformation of ~0.35 reflects the breaking of the structure.

Interestingly enough, the behavior of the system under Y-strain turned out to be completely different (Figs. 3, 4). In this case, the lattice tends to disintegrate into a set of single thread-like 1D BN molecules which form in this case an ordered orthorhombic structure. The associated behavior of the force and energy for the h-BN monolayer shows a structural transformation. Above the point of transformation, $\Delta y/y=0.3$, the tensile force changes its sign, indicating the onset of weak repulsion in the system.

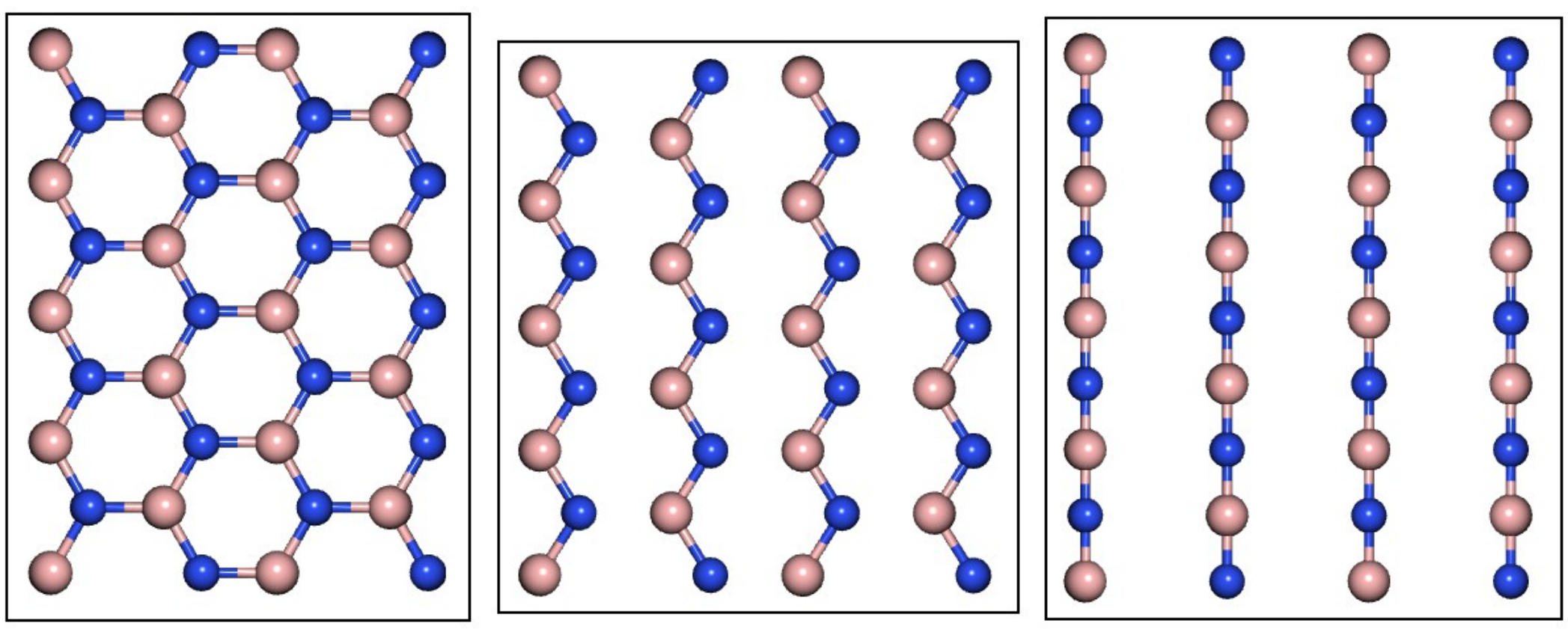

**Fig. 3** Structural changes occurring in 2D h-BN during the uniaxial deformation in armchair direction. Left: the initial h-BN lattice; center: relaxed structure at the strain $\Delta y/y = 0.2$; right: relaxed structure at the strain $\Delta y/y = 0.3$.

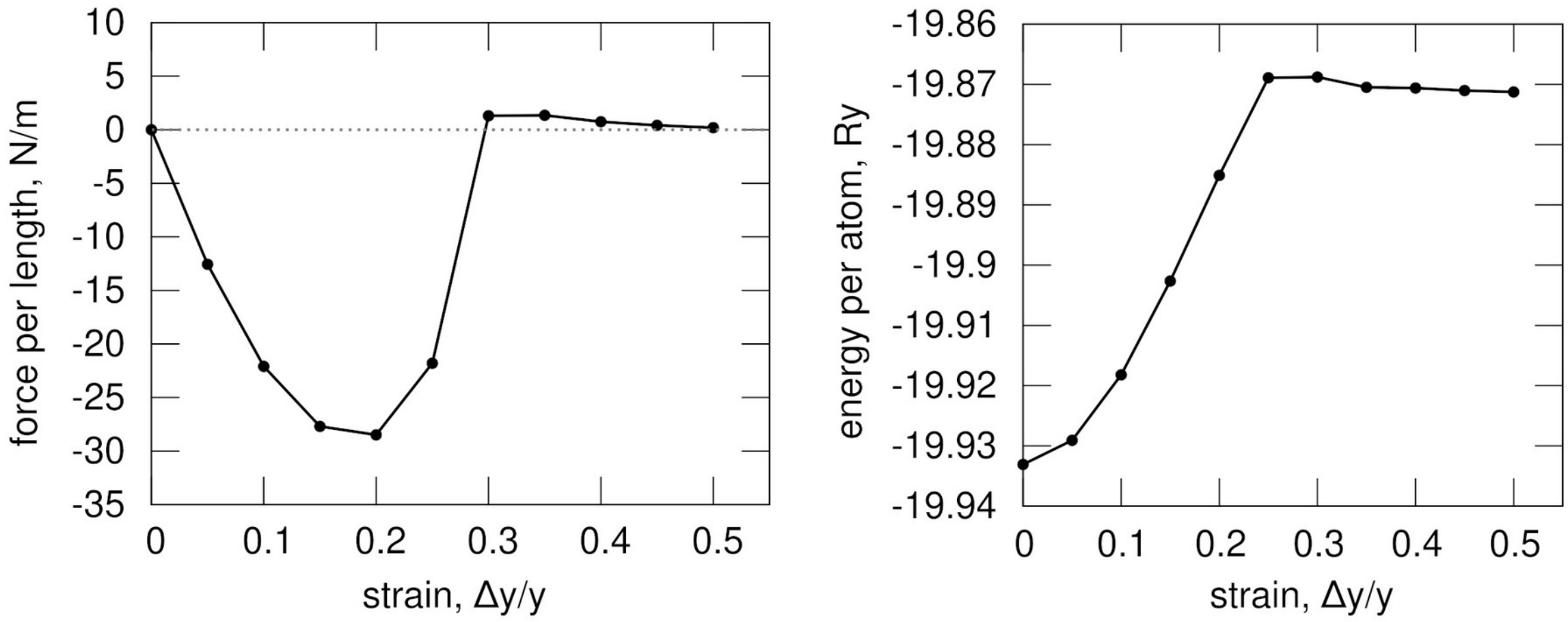

**Fig. 4** Calculated tensile force per length (left) and energy per atom (right) vs Y-strain for h-BN monolayer.

Thus, this 2D structure consisting of 1D thread-like molecules is clearly unstable. It is to mention that a closely related tetragonal (*I4/mmm*) bulk BN phase consisting of similar 1D threads has been theoretically predicted earlier (Jain et al. 2013).

Most interesting findings were done in the case of shear deformations. The simulations performed for the supercell containing 16 atoms revealed that the 2D h-BN lattice subjected to the combined fixed shear strain, Δy/x=0.35 and Δx/y = 0.25, undergoes a series of structural transformations in the course of relaxation, Fig. 5. The associated energy behavior given in Fig. 6 reflects the formation of respective transitional structures. All the configurations A, B, C, D are strictly two-dimensional (with zero thickness), while the final configuration with the lowest energy, D1 (not displayed in Fig. 5), is crimped in Z-dimension. At the same time, the structures D and D1 are topologically equivalent, and, when allowed to freely (including lattice parameters) relax, both converge to the same strictly 2D orthorhombic lattice, which we denote below as o-BN phase. This indicates, in particular, the relative stability of the latter.

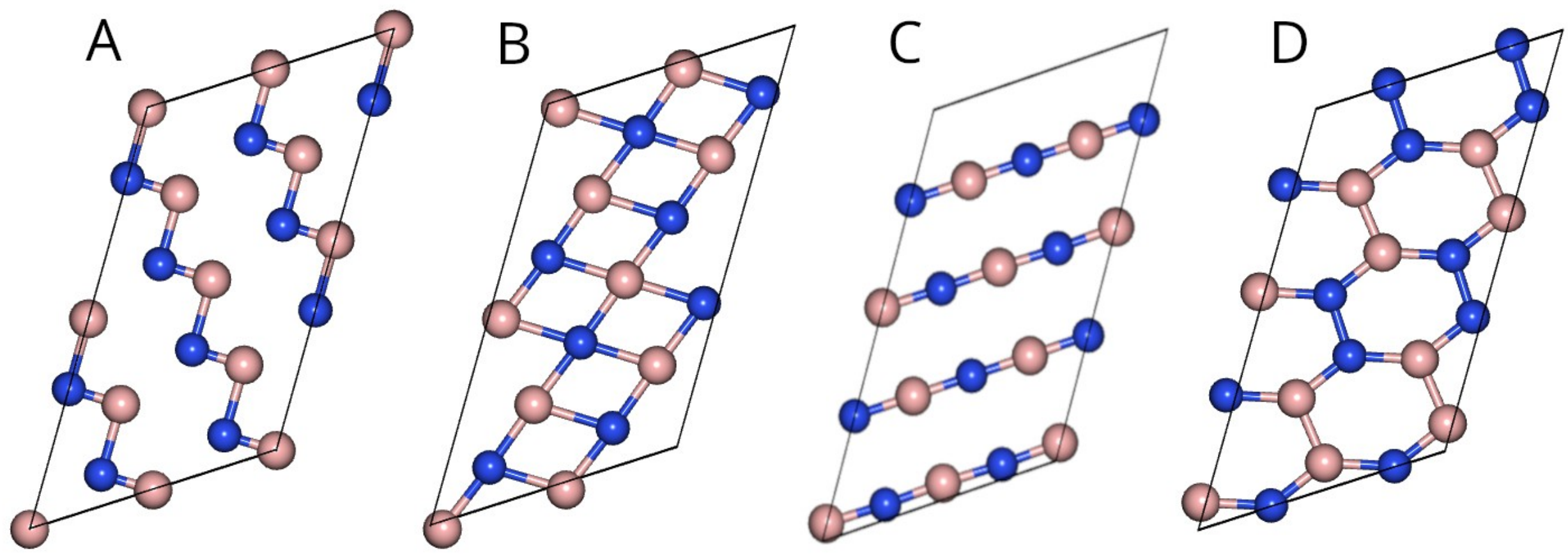

**Fig. 5** A series of structural transformations occurring in h-BN for the fixed shear strain set as  Δy/x=0.35 and Δx/y = 0.25. A) the initial configuration formed immediately after applying the strain; B), C), and D) – transitional structures formed in the process of ionic relaxation while keeping the cell strain fixed.

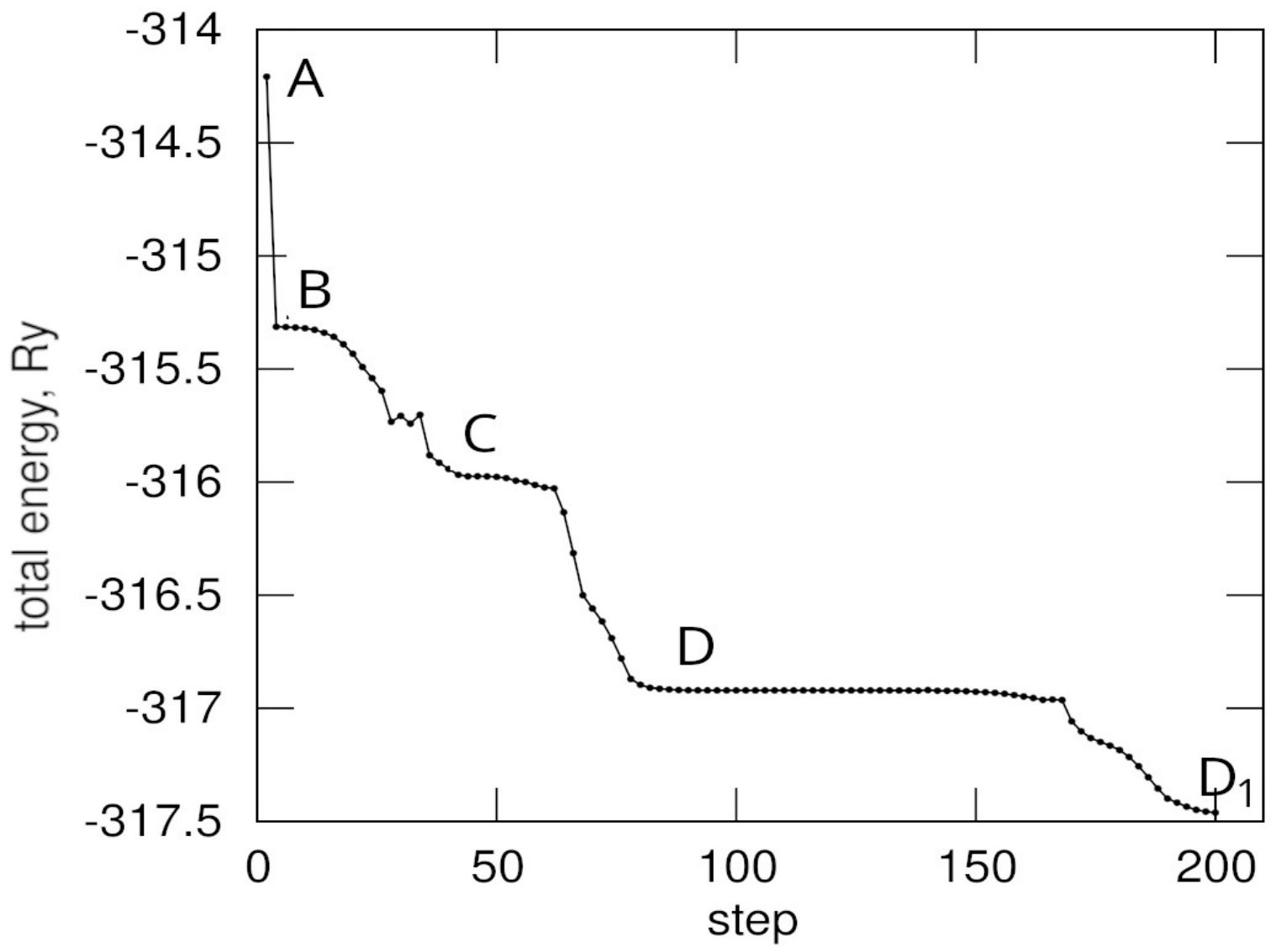

**Fig. 6** The energy behavior reflecting the structural transformations given in Fig. 5.

To have an idea about the properties of this orthorhombic phase, we calculated its basic structural and electronic characteristics. The examination performed with the use of PYMATGEN software (Ong et al. 2013), indicates that the unit cell of o-BN contains 4 atoms (Fig. 7, right), and has the 2D layer space symmetry *pmmm* (No. 37 in the respective international classification (Fu et al. 2024)). Calculation of the band structure of the BN compounds under consideration reveals that, in contrast to the conventional phase 2D h-BN, the orthorhombic phase o-BN has a narrow band gap ~ 0.65 eV typical for semiconductors (Fig. 8, right). In addition, we calculated the formation energies with respect to the stable bulk phase of boron, $\alpha$-$B_{12}$ (*R-3m*), and molecular nitrogen, $N_2$.

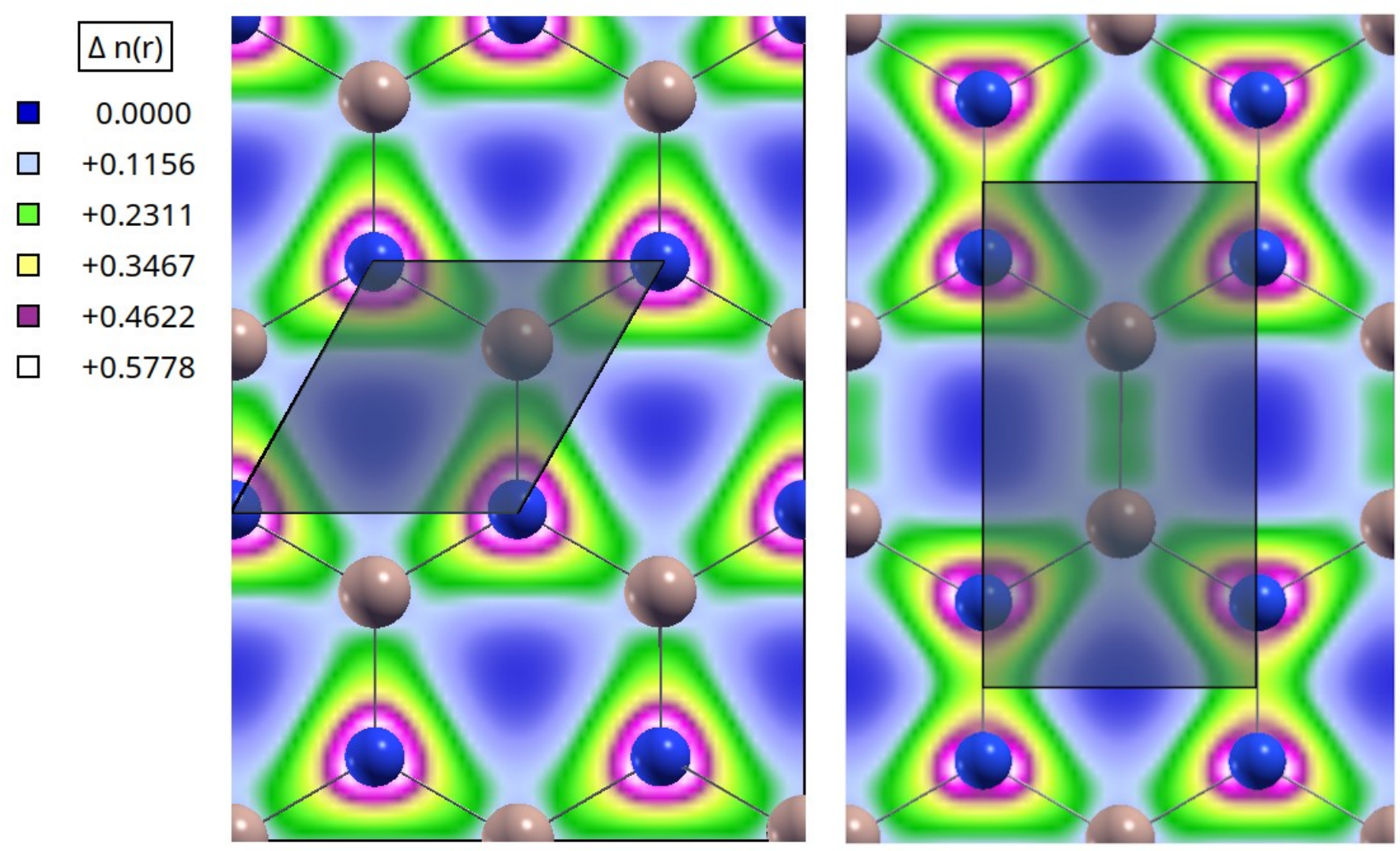

**Fig. 7** Comparison of the conventional monolayer h-BN lattice (*p-6m2*) (left), and the orthorhombic phase o-BN (*pmmm*) obtained in simulations (right). The unit cells are highlighted in gray. Along with the configurations, the spatial distribution of electron

densities is given.

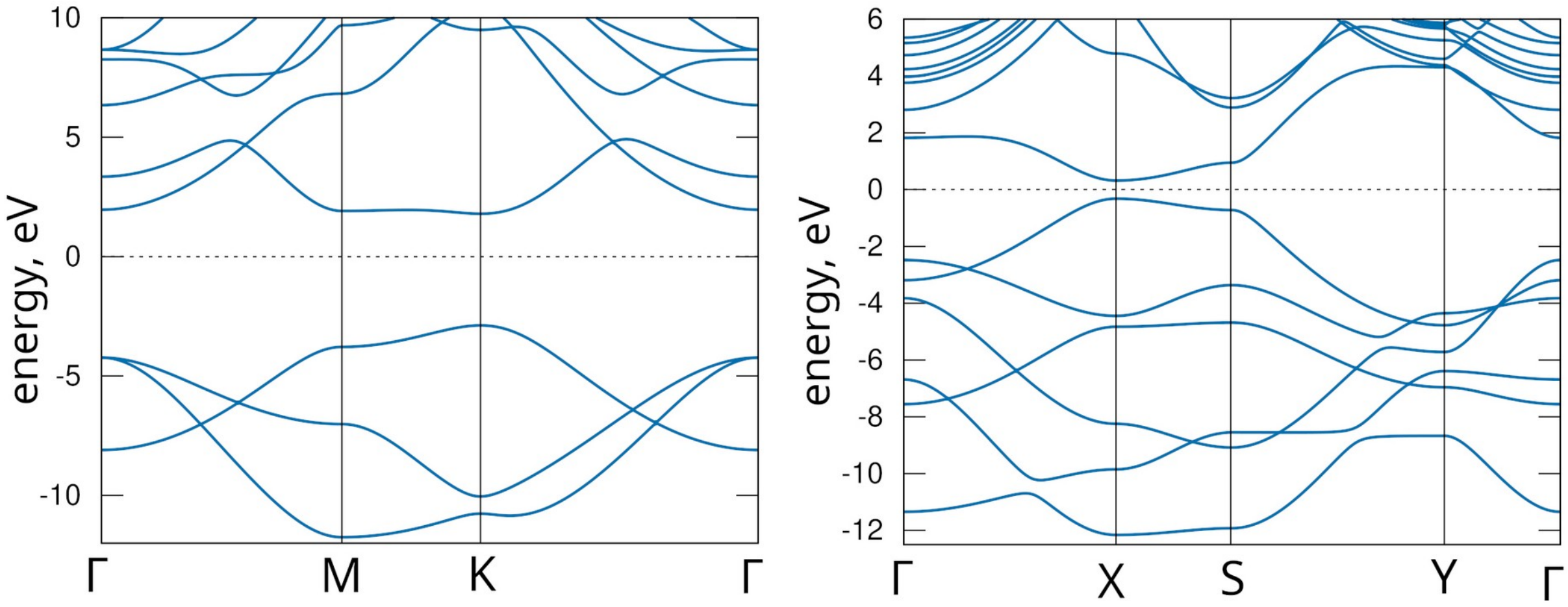

**Fig. 8** Comparison of electronic band structures calculated for the hexagonal h-BN (left) and orthorhombic o-BN 2D phases (right).

The comparison of the basic parameters of the conventional 2D h-BN and the predicted 2D o-BN phase is given in Figs. 7, 8 and Table 1.

**Table 1** Comparison of the parameters of the 2D h-BN and o-BN structures obtained in DFT calculations.

| 2D BN phase | Number of atoms in primitve cell | Crystal system, 2D symmetry | Cell parameters a, b (A), gamma (grad) | Density ($g/cm^2$) | Band gap (eV) | Formation energy (eV) |
|---|---|---|---|---|---|---|
| h-BN | 2 | Hexagonal p-6m2 [78] | a=b=2.51; γ=120 | $7.55 \cdot 10^{-8}$ | 4.7 | -1.27 |
| o-BN | 4 | Orthorhombic pmmm [37] | a=2.49; b=4.57; γ=90 | $7.24 \cdot 10^{-8}$ | 0.65 | -0.45 |

As is seen from Fig. 7, the orthorhombic o-BN structure differs in that it contains N-N bonds which accumulate considerable electron density. Such bonds are quite untypical for known bulk BN phases, which always tend to form B-N bonds only (see, for instance, the materials database (Jain et al. 2013)). This suggests, in particular, that the 2D orthorhombic phase under consideration cannot be obtained by conventional exfoliation procedure, since there is no appropriate parent bulk phase known.

An important issue concerning theoretically predicted phases is their mechanical and thermal stability. There are two well established theoretical approaches to address this issue, the examination of phonon properties and MD study. The examination of the phonon dispersion allows one to exclude the presence of imaginary frequencies, which are associated with the dynamical instability of phases, and MD simulations can directly demonstrate the stability of lattice parameters, energy and temperature of a dynamical system (see, for instance, (Bystrenko et al. 2024; Matar and Solozhenko 2023)).

We performed high precision calculation of phonon dispersion curves for 2D phase o-BN (*pmmm*) in harmonic approximation. The results clearly indicate the absence of unstable phonon modes, therefore, this theoretically predicted phase is expected to be dynamically stable at zero temperature (Fig. 9).

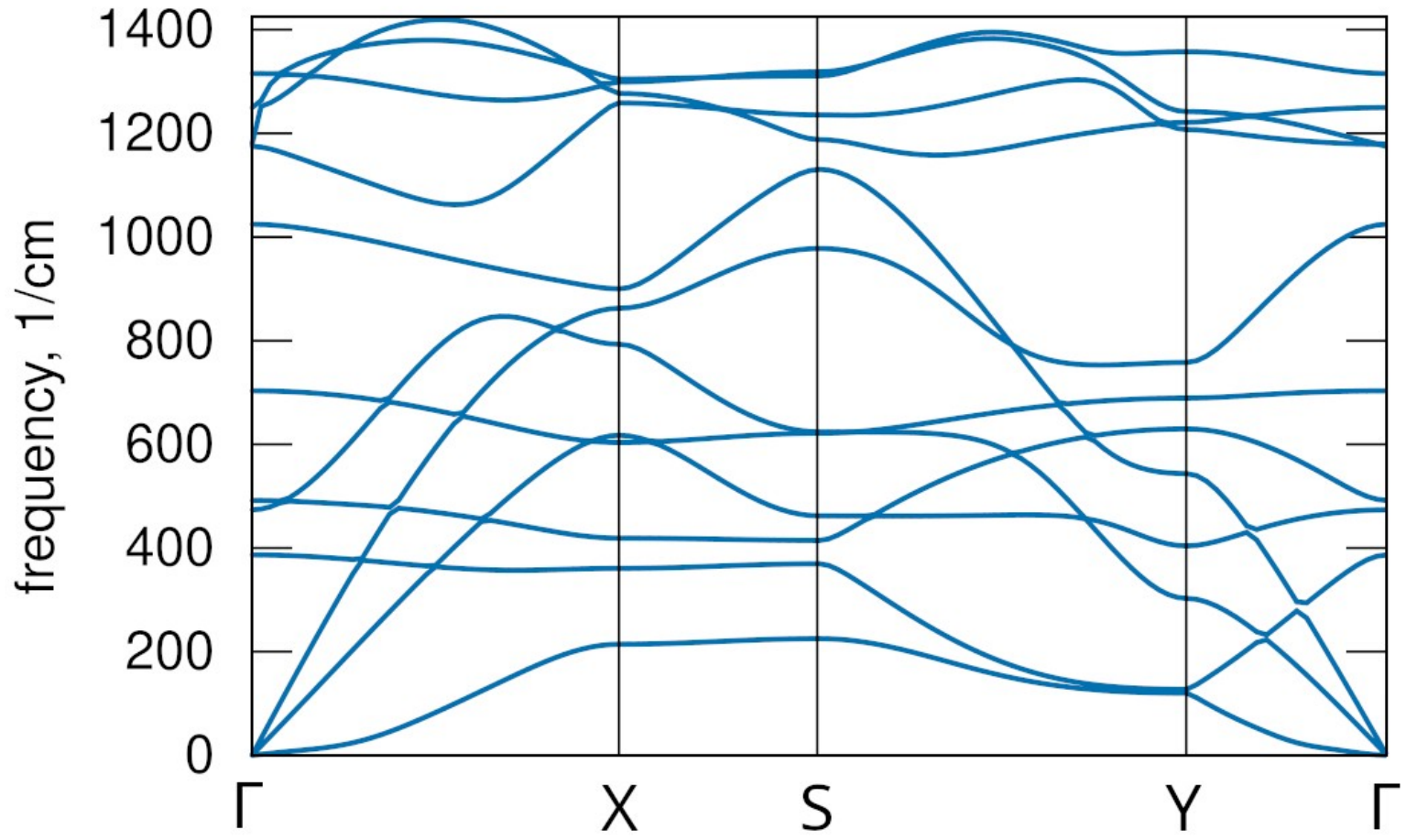

**Fig. 9** Phonon dispersion curves calculated for the 2D orthorhombic phase o-BN.

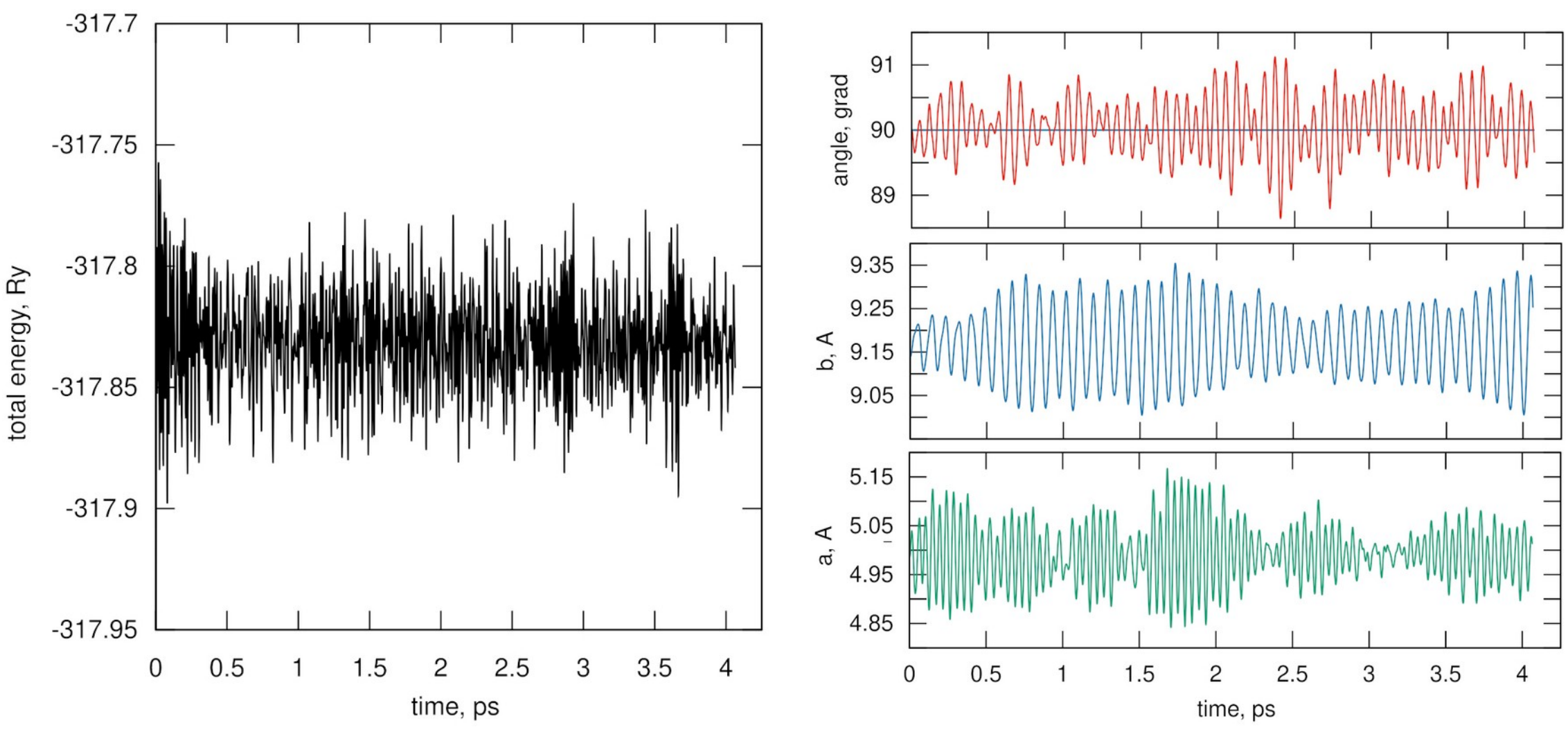

**Fig. 10** Behavior of the total energy (left) and cell parameters (right) of the 2D orthorhombic phase o-BN as obtained in Born-Oppenheimer MD simulations. The temperature of thermostat is maintained at ~ 870 K.

First-principle MD simulations were carried out within Born-Oppenheimer approximation by using the 'svr' (stochastic velocity rescaling (Bussi et al. 2007)) thermostat. We used a supercell containing 16 atoms, the time step was 0.25 fs and the total time span ~4 ps. The simulations performed for free geometry (i.e., by taking into dynamics lattice parameters) demonstrate the stability of basic characteristics of the system within the temperature range 300-900 K (Fig. 10).

Therefore, the phase o-BN remains thermally stable at these temperatures.

## IV Conclusions

To conclude, we examine structural transformations occurring in monolayer 2D hexagonal boron nitride subjected to strong deformations. The theoretical study is based on first-principle DFT approach. Depending on the direction and the type of strain, the simulations reveal different types of structural transformations in 2D h-BN associated with the elastic failure. It is shown that uniaxial tensile strain in armchair direction results in formation of unstable ordered structures consisting of 1D thread-like BN molecules. Shear deformations in certain directions lead to the emergence of a new structure, which, after full relaxation, can be identified as an orthorhombic o-BN lattice with 4 atoms in a unit cell and the 2D layer symmetry *pmmm*. This structure differs in that it contains N-N bonds, which are quite untypical for the conventional 2D and bulk BN compounds. Examination of phonon dispersion shows that the predicted phase o-BN is dynamically stable. The thermal stability within the temperature range of 300-900 K has been demonstrated by means of first-principle MD simulations based on Born-Oppenheimer approximation. The study of electronic structure indicates that o-BN phase is expected to be a semiconductor with the band gap ~ 0.65 eV. The obtained results demonstrate that the approach employed can be regarded, in particular, as an alternative to the conventional exfoliation procedure for predicting new 2D structures.

## Declarations

**Funding:** No funds, grants, or other support was received.

**Competing interests:** The authors have no competing interests to declare that are relevant to the content of this article.

**Data availability:** The data presented in this study are available upon request from the corresponding author.